\documentclass[twocolumn]{aastex631}

\usepackage{graphicx}

\usepackage{apjfonts}

\usepackage{enumitem}

\def\oiii {[\ion{O}{3}]}

\def\oiiiv {[\ion{O}{3}]$ ~\lambda 5007$}
\def\nii  {[\ion{N}{2}]}

\def\delv  {$\Delta \upsilon$}

\newcommand{\hb}{H{$\beta$}}

\newcommand{\kms}{${\rm km~ s^{-1}}$}
\newcommand{\sigs}{\ensuremath{\sigma_{\ast}}}

\def\sig{$\sigma$}

\newcommand{\dnfour}{$D_{n}(4000)$}
\newcommand{\hdela}{$\rm H\delta_{\rm A}$}

\newcommand{\sn}{{\it S/N}}

\newcommand{\lledd}{${\it L_{\rm{bol}}/L{\rm{_{Edd}}}}$}
\newcommand{\mgb}{\ion{Mg}{1}{$b$}}
\newcommand{\cahk}{\rm \ion{Ca}{2} H,K}

\newcommand{\lbol}{$L_{{\rm bol}}$}
\newcommand{\loiii}{\ensuremath{L_{\mathrm{[O~ {\tiny III}]}}}}

\received{$-----$}
\revised{$-----$}
\accepted{$----$}

\submitjournal{ApJ}

\shorttitle{\oiii\ profile comparisons}  
\shortauthors{Jin et al.}

\begin{document}

\title{Does Feedback from Supermassive Blackhole Co-evolve With Host In Type 2 Quasars?}
\correspondingauthor{}
\email{wj@nao.cas.cn; kmz@hebtu.edu.cn}
\author{S. Jin}
\affiliation{College of Physics, Hebei Normal University, Shijiazhuang 050024, People’s Republic of China}

\author{J. Wang}
\affiliation{Guangxi Key Laboratory for Relativistic Astrophysics, School of Physical Science and Technology, Guangxi University, Nanning 530004, China; wj@nao.cas.cn}
\affiliation{Key Laboratory of Space Astronomy and Technology, National Astronomical Observatories, Chinese Academy of Sciences, 20A Datun Road, Chaoyang District, Beijing 100012, China}

\author{M. Z. Kong}
\affil{College of Physics, Hebei Normal University, Shijiazhuang 050024, People’s Republic of China}

\author{R. J. Shen}
\affiliation{Purple Mountain Observatory and Key Laboratory of Radio Astronomy, Chinese Academy of Sciences, 10 Yuanhua Road, Nanjing 210033, China}
\affiliation{School of Astronomy and Space Science, University of Science and Technology of China, Hefei, Anhui 230026, China}

\author{Y. X. Zhang}
\affil{National Astronomical Observatories, Chinese Academy of Sciences, 20A Datun Road, Chaoyang District, Beijing 100012, China}

\author{X. D. Xu}
\affiliation{Key Laboratory of Space Astronomy and Technology, National Astronomical Observatories, Chinese Academy of Sciences, 20A Datun Road, Chaoyang District, Beijing 100012, China}
\affil{School of Astronomy and Space Science, University of Chinese Academy of Sciences, Beijing, China}

\author{J. Y. Wei}
\affiliation{Key Laboratory of Space Astronomy and Technology, National Astronomical Observatories, Chinese Academy of Sciences, 20A Datun Road, Chaoyang District, Beijing 100012, China}
\affiliation{School of Astronomy and Space Science, University of Chinese Academy of Sciences, Beijing, China}

\author{Z. Xie}
\affil{College of Physics, Hebei Normal University, Shijiazhuang 050024, People’s Republic of China}

\begin{abstract}
The feedback from accretion of central supermassive black holes (SMBHs) is a hot topic in the co-evolution of 
the SMBHs and their host galaxies. By tracing the large scale outflow by the line profile and bulk velocity shift of \oiii$\lambda5007$,
the evolutionary role of outflow is studied here on a large sample of 221 type 2 quasars (QSO2s) extracted from Reyes et al. 
 By following our previous study on local Seyfert 2 galaxies,
the current spectral analysis on the SDSS spectroscopic database 
enables us to arrive at following results: (1) by using  the Lick indices, we confirm that QSO2s are on average associated 
with younger stellar populations than Seyfert galaxies; (2) QSO2s with a stronger outflow are tend to be associated with a younger 
stellar population, which implies 
a coevolution between the feedback from SMBH and the host in QSO2s; (3) although an occupation at the high \lledd\ end,
the QSO2s follow the \lledd-\dnfour\ sequence established from local, less-luminous Seyfert galaxies, which suggests a decrease of accretion 
activity of SMBH and feedback as the circumnuclear stellar population continuously ages. \rm
\end{abstract}


\section{Introduction} \label{sec:intro}

There is accumulating evidence supporting a fact that feedback process from supermassive black holes (SMBHs)
plays an important role in the conception of co-evolution of growth of SMBHs and their host galaxies 
where the SMBHs reside in 
\citep[see reviews in][]{heckman04,Alexander12,Fabian12}. 
In both secular and merger evolutionary scenarios proposed in past decades \citep[e.g.,][]{Sanders1988,Di2008,Hopkins08,Hopkins09,Draper12,Shankar2012,Hopkins13,Heckman14}, 
the feedback is generally required in not only semianalytic  models but also in numerical
simulations to self-regulate growth of SMBH
and star formation occuring in the host galaxy by either suppressing star formation through sweeping out
circumnuclear gas or triggering star formation by compressing the gas  \citep[e.g.,][]{Alexander12,Page12,Kormendy13,Zubovas13,Ishibashi14,Cresci15,Carniani16,Villar16,Woo17,Cresci18,2020Perna,Scholtz21,Woo21};  
In fact, the models with feedback 
can reproduce the firmly established $M_{\mathrm{BH}}-\sigma_\star$ relation, luminosity functions
of both quasars and normal galaxies \citep[e.g.,][]{Haehnelt1998,1998Silk,1999Fabian,2000Kauffmann,Granato04,Springel05,Croton06,Di07,Hopkins08,Khalatyan08,Menci08,Somerville08}.
Furthermore, the ``over cooling'' problem in
the $\Lambda$ cold dark matter ($\Lambda$CDM) galaxy formation model
can be solved by an additional heating contributed by the feedback from 
SMBHs  \citep[e.g.,][]{Ciotti07,Somerville08,Hirschmann14}.

On the observational ground, the feedback
is usually traced by the frequently observed outflows from 
central SMBHs to study not only its origin but also its effect on host galaxies
 \citep[see reviews in][]{Veilleux05,Fabian12}.  
Among the different diagnosis of the outflows, the mostly used is the 
blue asymmetry of the prominent [\ion{O}{3}]$\lambda\lambda$4959, 5007
doublet and its bulk blueshift with respect to the local system, which 
is found to be quite prevalent in both local and distant active galactic nuclei 
\citep[AGNs,e.g.,][]{Heckman81,V01,Zamanov02,Marziani03,Aoki05,Boroson05,komossa08,Xu09,Vi11,Liu13,mullaney13,Zhang13,Harrison14,Vil14,Karouzos16,Wang16,wang18}.

Using the line profile of the [\ion{O}{3}] line emission,   
\citet {wang11} performed a comprehensive study on a large sample of local 
obscured AGNs, which is extracted from MPA/JHU value-added catalog \citep [e.g.,][] {Kauffmann2003}
based on the Sloan Digital Sky Survey \citep [e.g.,][] {York00}, to explore the 
relation between outflows and properties of the host galaxies, according to the 
widely accepted AGN's unification model \citep [e.g.,][] {Antonucci93} in which 
the central AGN's continuum and emission from the broad-line region (BLR) are obscured by the torus. 
The authors proposed a trend that the local Seyfert 2 galaxies with stronger blue
asymmetries tend to be associated with not only younger stellar populations, but 
also higher AGN Eddington ratio ($L/L_{\mathrm{Edd}}$, where $L_{\mathrm{Edd}}=1.26\times10^{38} M_{\mathrm{BH}}/M_\odot\ \mathrm{erg\ s^{-1}}$
is the Eddington luminosity). This result is further confirmed in \citet {wang15} for a sample of nearby partially obscured AGNs. 

A question is therefore naturally arisen: is the dependence of strength of outflow on 
both stellar population ages and $L/L_{\mathrm{Edd}}$ revealed in local Seyfert galaxies still valid for 
their either high luminosity or high redshift cousin? This question is motivated from two 
facts. On the one hand, there is ample evidence supporting that the growth of small SMBHs in low luminosity AGNs (e.g., Seyfert galaxies)
is dominantly through a secular evolution in which the gas inflow towards a central SMBH
is mainly caused by an instability or viscous of the gas \citep [e.g.,][and references therein] {Heckman14,Wang16,Wang19} .       
On the contrary, a gas-rich major merger is preferred for the luminous quasars with a large SMBH as implied by the studies 
of the host galaxies of quasars and ultra-luminous infrared galaxies \citep [ULIRGs, e.g.,][] {Sander88,Bahcall97,Kirhakos99,Hao05,hou11}.  
On the other hand, the cosmic co-evolution of SHBM growth and star formation 
implies that the feedback from AGNs is strong in early universe when the peaks of both AGN activity and
star formation occur roughly coincident \citep [e.g.,][] {hou11,Ishibashi13,Harrison17} .

In this paper, we try to answer the aforementioned question by extending the study in \citet {wang11} to 
type 2 quasars (QSO2s), the high luminosity counterparts of the local Seyfert 2 galaxies.  
In fact, previous studies through different methods indicate that the outflow phenomenon is quite popular in QSO2s
\citep [e.g.,][] {Vi11,Liu13,Vil14,Karouzos16,Almeida17}.
The scale of the violent outflow in QSOs ranges from a few pc ($\upsilon\sim0.1c$) to $\sim$kpc 
($\upsilon\sim10^{2-3}\ \mathrm{km\ s^{-1}}$) 
\citep [e.g.,][]{Pounds03,Harrison16,Schreiber19,Kakkad20}. An outflow with an extension of 
13pc from the center SMBH has been identified for the ionized gas in 
the obscured AGN XID\,2028 at $z=1.59$ \citep{Carniani16}, which is 
recently confirmed by the Early Release Science JWST NIRSpec observations
\citep [e.g.,][] {Cresci23}. Compared with the VLA 3 GHz map, the authors argue that the extended outflow in the object is likely related with the low-luminosity radio jet.
\rm

The outline of this paper is as follows. The sample selection and
spectral analysis are described in Sections 2 and 3, respectively.
Section 4 shows the statistical results. A discussion is presented in 
Section 5. \rm 
\rm
A  $\Lambda$CDM cosmology with
parameters $H_0=70\mathrm{km\ s^{-1}\ Mpc^{-1}}$, $\Omega_m=0.3$, and $\Omega_\Lambda=0.7$
\citep{Spergel03} is used throughout the paper.

\section{Sample selection} \label{sec:sample}

A couple of QSO2s samples have been published in past decades \citep[e.g.,][]{zaka03,reyes08,yuan16},  thanks to the SDSS survey.
We start from the QSO2s sample provided by \citet{reyes08}, simply because  
\citet{kong18}  performed a comprehensive study on their black hole mass and \lledd\ by adopting 
the narrow emission lines, such as \oiiiv\, as an indicator.

The used QSO2s catalog contains in total 887 objects with redshift  $z < 0.83$, 
whose \oiiiv\ line luminosity  ranges from  $10^{8.3} L_{\odot}$ to $10^{10.0}L_{\odot}$.
They are selected from the seventh data release of SDSS \citep{abaz09} 
by following the four main selection rules (please see selection details in \citealt{zaka03, reyes08}):
\begin{enumerate}
\item
In  order to retain objects with weak continuum and strong narrow \oiii$\lambda$5007 emission lines,
its rest-frame equivalent width
 is  greater than 4 and
the corresponding signal-to-noise ratio ($\rm S/N$)  is  $\geq 7.5$ 
over the entire spectroscopic range of 3800-9200 \AA \rm.
 \item
For objects with $z < 0.36$,   H$\beta$+\oiii$\lambda\lambda$4959,5007, H$\alpha$+\nii$\lambda\lambda$ 6548,6583 and 
[\ion{S}{2}]$\lambda\lambda$6716, 6731 lines are 
required to be available.  AGNs are distinguished  from star-forming galaxies by using the line ratio diagnostic criteria presented by \citet{kewley01},
i.e., $\log\left(\cal{[\rm{OIII}]/H\beta}\right)>\frac{0.61}{\log({\rm [NII]6583/H\alpha})-0.47}+1.19$, and $\log\left(\cal{[\rm{OIII}]/H\beta}\right)>\frac{0.72}{\log({\rm [SII]/H\alpha})-0.32}+1.30$.

 \item
For objects with redshift $0.36 \leq z < 0.83$, the \hb\ line is required to either be undetected or have
a line ratio of $\rm log \frac{[\rm{OIII}]~5007}{H\beta} > 0.3$ if the flux of the \hb\ line is  available with $ \sn\ > 3$.

\item
Moreover, for objects with $z >0.6$, the FWHM of \ion{Mg}{2}$\lambda$2800 emission line is required to be less than 2000 \kms\ if
the line is above the noise level. 
\end{enumerate}

The SDSS spectra of the 887 QSO2s candidates were then checked visually one-by-one by us. 
More than 40 objects with unambiguous double-peaked line profile were excluded to 
avoid unreliable \oiii\ $\lambda5007$ emission-line profile measurements. The double-peaked 
profiles may be caused by galaxy mergers, disk rotation in large scale, or bipolar outflows (e.g., \citealt{liux10, ge12}).  
Objects with either \oiii\ line \sn\ less than 10 or incomplete line profile were additionally removed. 
Finally, there are 772 QSO2s for subsequent spectral analysis. 

\section{Spectral Analysis} \label{specfit}

For each of the objects listed in the sample,
the SDSS spectrum is at first corrected for Galactic extinction by 
using the extinction curve of \citet{cardelli89}  and the extinction value adopted from \citet{schlafly11}.
The redshift provided by the SDSS pipelines is then used to transform the observed spectrum to the rest-frame.

\subsection {Continuum and Stellar Subtraction}
Before the \oiii\  emission line measurements, the underlying stellar component including 
stellar continuum and stellar absorption features must be carefully substracted.
This substraction was  carried out by the  publicly automatic  procedure  of 
the penalized pixel-fitting ({\tt pPXF}) \citep{cappellari04,cappellari17}  method.
We refer the readers to \citet{kong18} for the details of the stellar component
subtraction for this sample and stress here some key issues as follows:
\begin{enumerate}
\item
The spectral fitting region covers 4100--5400 \AA\ including \hb, \oiiiv\ emission line.
\item
The Indo-U.S. stellar spectral library \citep{valdes04} is adapted for fitting the continuum of the host galaxies.
The stellar spectral library  has a spectral resolution of FWHM = 1.35\AA\ and with a wavelength range 3460--9464 \AA.
\item
The  IDL package {\tt mpfit} \citep{markwardt09} is used to 
determine the best fitted parameters through a $\chi^{2}$ minimization in which the nonlinear Levenberg-Marquardt algorithm is adopted.
\item
In addition to the strong emission lines (such as \ion{He}{2} $\lambda 4686$, 
H$\beta$, and \oiii\ $\lambda\lambda 4959$, \ 5007),  the 
H$\gamma$ $\lambda4340$, {\oiii\ $\lambda 4363$, [\ion{O}{2}]$\lambda 3727$, 
[\ion{Ne}{3}]$\lambda3869$ and \mgb\ $\lambda\lambda5167$,\ 5173,\ 5184 }triplets are excluded from the  the fitting.
The \mgb\ triplets  were  also masked because of their potential systematic effects caused by [Mg/Fe] enhancement. 
\end{enumerate}

\begin{figure*}[ht!]
\centering
\epsscale{1.2}
\plotone{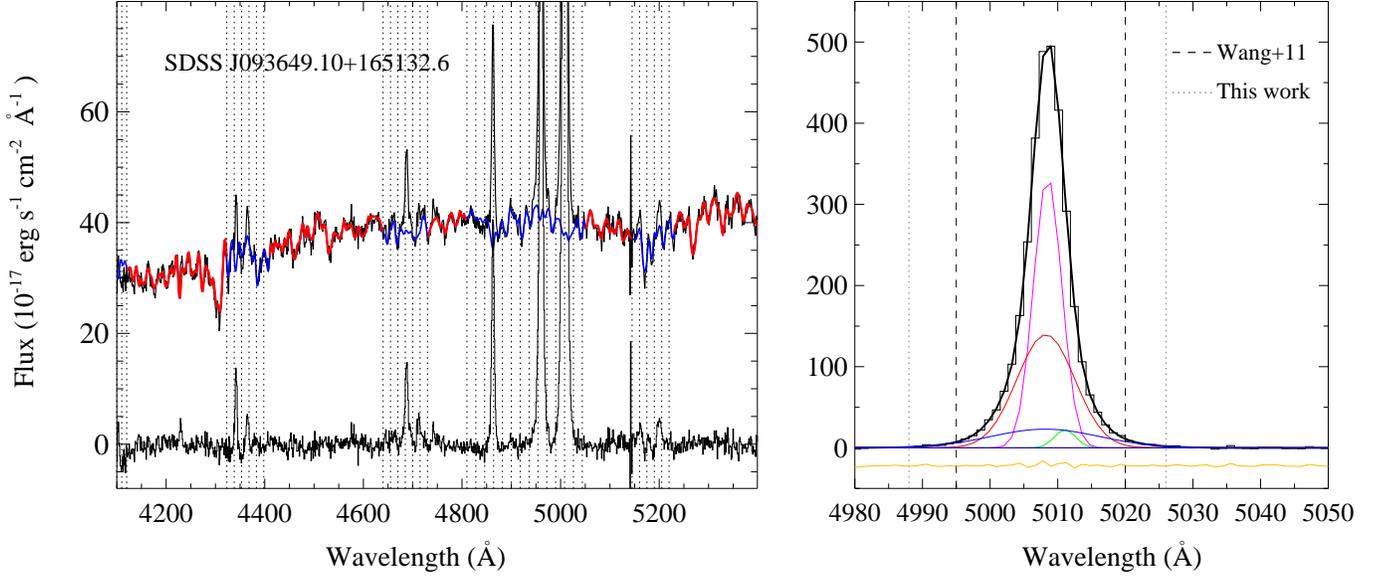}
\caption{\it Left: \rm An example of continuum/stellar subtraction.  The original and continuum-removed spectra are displayed 
by the upper and lower black curves, respectively. The best-fitted stellar component is overplotted with the red and blue colors for 
the fitting region and the region excluded in the fitting, respectively. \it Right: \rm  The \oiii$\lambda5007$line profile modeled by a linear combination of set of
Gaussian components, after a removal of the continuum. The observed and modeled line profiles are plotted with thin and heavy 
black solid lines, respectively. Each Gaussian component is plotted with a thin colorful line. The orange curve below the line spectrum presents the residuals 
between the observed and modeled emission line profiles. The vertical long and short dashed lines mark the wavelength regions used for measuring line profile parameters in 
\citet{wang11} and in this work, respectively.
\label{fig:constar}}
\end{figure*}

An example of continuum and stellar fitting can be seen from the left panel in Figure \ref{fig:constar}.

\subsection{Measurements of emission-line profile parameters}
After the continuum and  absorption stellar components are subtracted from each spectrum, a set of shape parameters are measured for the \oiii line 
profile in this section\footnote{ As illustrated in Figure 2, the 
continuum removal is necessary before an emission-line analysis in 
type-II AGNs, because the measured \oiii\ 
line profile asymmetry strongly depends on the behavior of line wing and 
the determined continuum level, which can be distorted heavily by the 
absorption features of the starlight component.}.

\subsubsection{Line profile fitting}
Many ways are used to parameterize emission-line profile.
Parameters included the FWHM and the second moment of the line are commonly used.  The second moment is defined to be

\begin{equation}
\sigma^2=\bigg(\frac{c}{\lambda_{c}}\bigg)^2\frac{\int(\lambda-\lambda_{c})^2f_\lambda d\lambda}{\int\!f_\lambda d\lambda}
\label{eq:sigma2}
\end{equation}  
where $f_\lambda$ is the flux density of the continuum-subtracted spectrum, 
$\lambda_{c}$ is the line centroid and is defined as $\overline{\lambda}=\int\!\lambda f_\lambda d\lambda/\int\!f_\lambda d\lambda$.
Both two parameters comparably describe the line broadening for a pure Gaussian profile, i.e.,
$\mathrm{FWHM}=2\sqrt{2\ln2}\sigma\approx2.35\sigma$.
As described in \citet{greeney05},  \sig\ is more sensitive to the  line wings and becomes relatively broader, which indicates that \sig\
contains more information on the line profile broadening if the profile is not
 a pure Gaussian profile. In fact, in addition to \sig\, 
as presented in \citet{binney98},  
$\xi_3$ and $\xi_4$, which is the high-order dimensionless line shape parameters, can be used to parameterize line profile deviation from a pure Gaussian profile.

$\xi_k$ is defined as 

\begin{equation}
\xi_k=\mu_k/\sigma^k\ k\geq3
\label{eq:xi}
\end{equation} 
where \sig\ is the second-moment defined above, and 
$\mu_k$ the $k$-order moment defined below
\begin{equation}
\mu_k=\bigg(\frac{c}{\overline{\lambda}}\bigg)^k\int (\lambda-\overline{\lambda})^kf_\lambda d\lambda
\label{eq:miuk}
\end{equation}

$\xi_3$, the so-called ``skewness'', measures a deviation from a symmetric profile.
$\xi_3=0$ corresponds to a symmetry profile. Meanwhile,
$\xi_3>0$ denotes a red asymmetry, and $\xi_3 < 0$ a blue asymmetry.

$\xi_4$, the so-called ``kurtosis'', measures a symmetric deviation from 
a pure Gaussian profile with $\xi_4=3$.
$\xi_4>3$ corresponds to an peaked emission line profile superposed on a 
broad base, and  $\xi_4<3$ to a emission line profile like a ``boxy'' shape.
We refer the readers to Figure 11.5 in 
\citet{binney98} for how the line shapes change with the values of $\xi_3$ and $\xi_4$ in details.

The \oiii\ $\lambda5007$ blue wings are found to overlap with the \oiii\ $\lambda4959$ lines in some objects with strong outflows.    
In order to avoid a distortion due to this overlapping, we first fit each \oiii\ doublet with a linear combination of a set of Gaussian components, in which 
the doublets have the same width and fixed line flux ratio of
the theoretical value of 1:3. 
An example of the profile modeling is shown in the right panel of Figure 2.
The \oiii\ $\lambda$5007 line profile parameters are then measured from the modeled line profile.
 The wavelength range over which the line profile parameters are measured should be carefully choiced.  
 The wavelength range of 4995-5020\AA\AA\ was used for measuring $\xi_3$ and $\xi_4$ for Seyfert 2 galaxies sample in  \citet {wang11}. However, 
 this wavelength range can not cover most of the  \oiii\  profiles well for the type  2 quasar (QSO2s) sample, because
 their  \oiii\  line profiles usually have relatively broader line wings.
We therefore instead measure their parameters within the wavelength range where the  line flux level 
is 2 times of the continuum fluctuation that is assessed in the wavelength range free of strong emission or absorption lines, i.e.,  between 4400--4600\AA\AA.
In Section \ref{sec:erroranaly}, we discuss the  wavelength range effect on the parameters measurements.
 
 \subsubsection{Instrumental resolution}
 
The observed \sig\ is resulted from a convolution of the true line profile and the instrumental profile.
By assuming the profiles can be described as a pure Gaussian function, 
the intrinsic line width \sig\ can be estimated approximately by 
 $\sigma^2=\sigma^2_{\mathrm{obs}}-\sigma^2_{\mathrm{inst}}$, where $\sigma_{\mathrm{obs}}$ and $\sigma_{\mathrm{inst}}$ are the observed line width 
and the instrumental resolution, respectively. 
However, as stated in the Appendix of \citet{wang11}, the correction of the instrumental resolution is not a easy task 
for a non-Gaussian line profile, where  the  correction depends on the amount of deviation from a pure Gaussian profile. 
In their work, only the \oiii\ line width greater than $2\sigma_{\mathrm{inst}}$ were kept for subsequent analysis.
By following this sample selection rule, there are only six objects with $\sigma_{\mathrm{obs}} < 2\sigma_{\mathrm{inst}}$,
and 767 objects are left.

\subsubsection{\texorpdfstring{\oiii$\lambda5007$\ }.line relative velocity shifts}
Given the measured line centroids $\overline{\lambda}$,
we measure the bulk velocity shift of \oiii\ emission line $\Delta\upsilon=\delta\lambda/\lambda_0 c$ in each
stellar-subtracted emission-line spectrum, where $\delta\lambda$ is
the \oiii\ line wavelength shift relative to the galaxy rest frame determined from the absorption features of 
the host galaxy, 
$\lambda_0$ 
the wavelength of the [\ion{O}{3}] emission line in the rest frame, and $c$ the light velocity. A positive value of $\Delta\upsilon$ denotes a red bulk 
velocity shift, and a negative one a blue shift. \rm

\subsubsection{\texorpdfstring{\dnfour}. and \texorpdfstring{\hdela}.}

\begin{figure}[ht!]
\centering
\epsscale{1.18}
\plotone{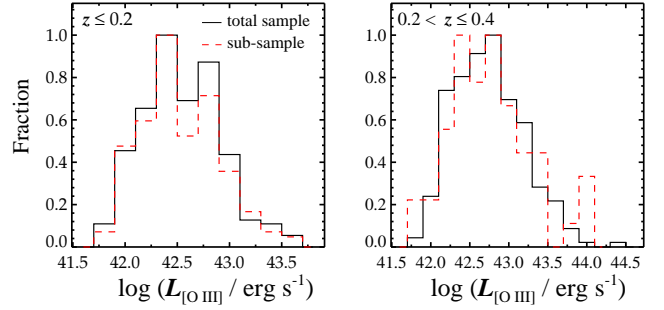}
\caption{ A comparison of the distributions of line luminosity of \oiii$\ \lambda5007$ in two redshift bins between the 221 QSOs used for subsequent statistic study and the parent QSO2s sample
given in  \citet{reyes08}. 
\label{fig:zlo3}}
\end{figure}

The two Lick indices of  \dnfour (the 4000\AA\ break) and \hdela (the equivalent width of the H$\rm \delta_A$ absorption 
feature of A-type stars) are good age indicators of the stellar populations of galaxies \citep[e.g.,][]{bruzual83,worthey97,balogh99}. 
The 4000\AA\ break is defined as $D_n(4000)=\int_{4000}^{4100}f_\lambda d\lambda/\int_{3850}^{3950}f_\lambda d\lambda$.
The index H$\delta_{\rm{A}}$ is defined as 
$\mathrm{H}\delta_A=(4122.25-4083.50)(1-F_I/F_C)$, where
 $F_I$ is the flux within the feature bandpass of $\lambda\lambda4083.50-4122.25$, and $F_C$ is the flux of
the pseudo-continuum assessed from the two beside regions: blue $\lambda\lambda4041.60-4079.75$ and 
red $\lambda\lambda4128.50-4161.00$. In order to avoid distortion by unreliable values, the \dnfour\ is 
measured for only 221 objects with not only a continuum median \sn$>5$, but also obvious stellar features, 
such as  absorption lines of \cahk\ $\lambda\lambda~3934$,3968, after an  one-by-one visual inspection.  For each of the 221 objects, we measure 
 both \dnfour\ and \hdela\ 
from the modeled starlight component, rather than the original spectrum,
which suggests that the contamination caused by the [\ion{Ne}{3}]$\lambda3869$ emission line is negligible for the resulted \dnfour. 
In addition, the adopted  
continuum \sn\ requirement leads to a bias against QSO2s at a high redshift.
The redshifts of the 221 objects with measured \dnfour\ actually range from 
0.05 to 0.35, which shows overlaps with a fraction of the Seyfert galaxies with high redshifts (see Section 4.4 and Figure 8 for the details).
The distributions of the \oiii\ $\lambda5007$ line luminosity of the 221 QSO2s are shown and compared with those of the total sample 
in two redshift bins in Figure \ref{fig:zlo3}.

\subsubsection{Wavelength range effect on  the \texorpdfstring {$\xi_3$}. and \texorpdfstring{$\xi_4$}.}\label{sec:erroranaly}

Before subsequent statistical study,
we first study the systematic of  $\xi_3$ and $\xi_4$ due to different  wavelength ranges where the two parameters are assessed. 
Based on the continuum flux fluctuation estimated between 4400--4600 \rm\AA, 
both parameters are measured in three different wavelength ranges in which the specific line flux is above a base at different significance levels
$\sn_{\rm Min}=1$,\,2,\,3, where $\sn_{\rm Min}$ is defined as the ratio between the level of the base and the  continuum flux fluctuation. 

\begin{figure*}[ht!]
	\centering 
	\epsscale{0.99}
	\plotone{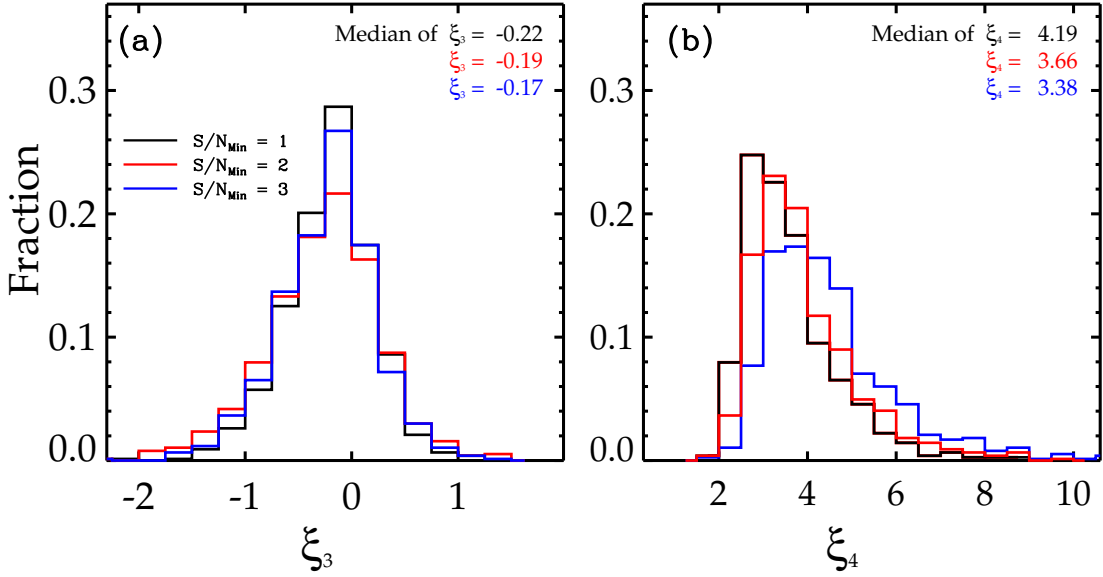}
	\caption{Distributions of  $\xi_3$ (left panel) and  $\xi_4$ (right panel) when they were measured at different 
	 wavelength ranges (or $\rm \sn_{\rm Min}$, see main text for the definition of  $\rm \sn_{\rm Min}$).
	 		\label{fig:it34comp}}
\end{figure*}

We argue that the systematics  on $\xi_3$ due to the adopted wavelength range is negligible, although it is not true for $\xi_4$.
Figure  \ref{fig:it34comp} compares the distributions of $\xi_3$ and $\xi_4$ measured within different wavelength ranges (or $\sn_{\rm Min}$).
The corresponding median values are compared in Column (2) in Table 1.
Columns (3)-(4) tabulate the matrix of two-sided Kolmogorov–Smirnov (K-S) tests, in 
which each entry denotes the maximum distance between the two distributions,
along with the corresponding probability that the two distributions come from 
the same parent sample shown in the bracket. 
The tests show that there is no significant difference between the three distributions, although larger the wavelength range used, slightly higher the median of $\xi_3$ will be.
However, with the increasing wavelength range, the distributions of $\xi_4$ clearly become wide, along with an increasing median value. 
The same tests yield a probability that the distributions are from the same parent sample as low as ${<10^{-6}}$.

As an additional test,  we measure both $\xi_3$ and $\xi_4$ by extending the wavelength range to 4000 \kms\ (about 60 \AA), which is 
twice the critical value that distinguishes the broad and narrow emission lines of AGNs. By comparing the values 
obtained with $\sn_{\rm Min}=1$,  the change of the median $\xi_3$ is less than 1\%, but the median of $\xi_4$  changes from 4.19 to 5.08.
This test again verifies the above statement that the adopted wavelength range has small (large) effect on measurements
of $\xi_3$ ($\xi_4$).

\begin{deluxetable}{ccccc}
	\tablecaption{Median and two-sided Kolmogorov–Smirnov test matrix of the  \oiii\ line shape parameters $\xi_3$ and $\xi_4$ measured within the different wavelength ranges.
		\label{tab:paraxi3}
	}
	\tablewidth{0pt}
	\renewcommand{\arraystretch}{1.5}
	\tablehead{
		\colhead{$\sn_{\rm Min}$} & \colhead{Median} & \colhead{1}& \colhead{2} & \colhead{3} \\
        \colhead{(1)} & \colhead{(2)} & \colhead{(3)} & \colhead{(4)} & \colhead{(5)}
	}
	\startdata
    \multicolumn{5}{c}{$\xi_3$}\\
    \hline
    1 & $-0.22\pm0.02$   & $\cdots $ & $\cdots $ & $\cdots$ \\
	2 & $-0.19\pm0.02$   & 0.051\ (0.267)  & $\cdots $ & $\cdots$ \\
	3 & $-0.17\pm0.01$   & 0.092\ (0.003) & 0.052\ (0.241) & $\cdots$ \\
    \hline
    \multicolumn{5}{c}{$\xi_4$}\\
    \hline
	1  &  $4.19\pm0.05$ & $\cdots $ & $\cdots $ & $\cdots $\\
	2  &  $3.66\pm0.04$ & 0.21\ ($< 10^{-15}$)  & $\cdots $ & $\cdots$\\
	3  &  $3.38\pm0.03$ & 0.32\ ($< 10^{-35} $) & 0.14\ ($< 10^{-7} $) & $\cdots $ \\
	\enddata

\end{deluxetable}

Finally, without further statement, the values of both $\xi_3$ and $\xi_4$ measured under the condition of 
$\sn_{\rm Min}=2$ are adopted for subsequent statistical studies, taking into account of a comparison with our previous studies in \citet{wang11} 
that was based on the observed spectra rather than the modeled line spectral profile.

\section{Results} \label{sec:result}

The measured parameters of the QSO2s are compared with our previous studies on Seyfert 2 galaxies in this section.

\subsection{Statistic of line Shape Parameters \texorpdfstring{$\xi_3$}. and \texorpdfstring{$\xi_4$}.}

\begin{figure}[ht!]
	\centering
	\epsscale{1.15}
	\plotone{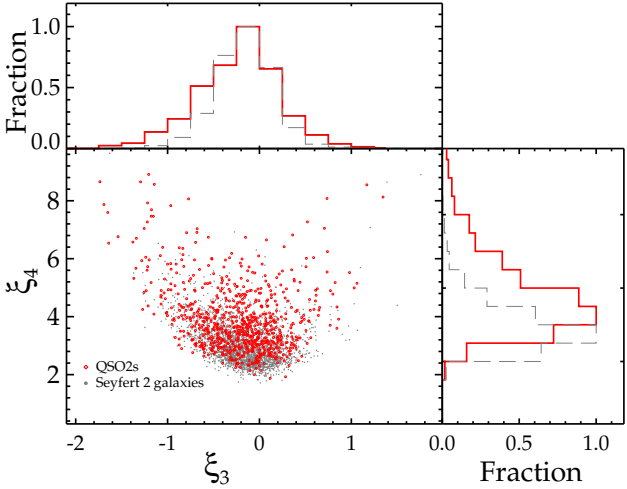}
	\caption{\it Main panel: \rm A comparison between the QSO2s (red points) studied here and Seyfert 2 galaxies (grey points) quoted from Wang et al. (2011) in the $\xi_3$ versus. $\xi_4$ 
		diagram. \it Upper left panel: \rm distributions
		of the parameter $\xi_3$ for the QSO2s (red line) and Seyfert 2 galaxies (grey line).
		\it Bottom right panel: \rm the same as the upper left panel but for the parameter $\xi_4$. 
		There are in total 767 objects with $\sigma_{\mathrm{obs}} > 2\sigma_{\mathrm{inst}}$ used in the plot.
		\label{fig:it34}}
\end{figure}

The occupations in the $\xi_3$ versus $\xi_4$ diagram are compared between  QSO2s and Seyfert 2 galaxies in Figure  \ref{fig:it34}.

The main panel in
the figure shows that both QSO2s and Seyfert 2 galaxies  
form a sequence starting from the pure Gaussian region (i.e., $\xi_3=0$ and 
$\xi_4=3$) to the upper left corner. Larger the blue asymmetry of the [\ion{O}{3}] line profile,
more peaked profile will be identified, which is consistent with the fact that two or more 
Gaussian components are usually required to properly reproduce both narrow core and blue wing 
of the observed [\ion{O}{3}] line profiles.

The distributions of $\xi_3$ and $\xi_4$ are presented in the upper and right sub-panels in Figure \ref{fig:it34}, 
respectively. Compared to the Seyfert 2 galaxies, the QSO2s tend to have stronger [\ion{O}{3}] 
blue asymmetry. The median value of $\xi_3$ is $-0.19\pm 0.02$ for the QSO2s, and $-0.16\pm 0.01 $ for 
the Seyfert 2 galaxies. A two-sided K-S test yields a difference between the two distributions 
at a significance level of $< 10^{-14}$ with a maximum absolute discrepancy of 0.16. 
A significant difference can be found for the $\xi_4$ distributions, in which 
the median value of $\xi_4$ is $3.66 \pm 0.04$ for the QSO2s, and 
$2.84 \pm  0.01 $ for 
the Seyfert 2 galaxies. Again, the same K-S test yields a difference at a 
significance level of $< 10^{-6}$ with a maximum absolute discrepancy of 0.39. In fact, this significant
discrepancy is not hard to be understood according to the fact that QSO2s typically have
stronger \oiii\ emission (mostly contributed by the narrow line core) than local Seyfert 2 galaxies.  
The large $\xi_4$ also indicates that QSO2s have stronger \oiii\ wings than the Seyfert 2 
galaxies.

\subsection{Dependence of line profile on stellar population}

The evolution of the  [\ion{O}{3}] line profile is examined in this section by using the two Lick indices \dnfour\ and \hdela,
which are widely used as age indicators of the circumnuclear stellar populations.

\begin{figure*}[ht!]
        \epsscale{0.9}
	\plotone{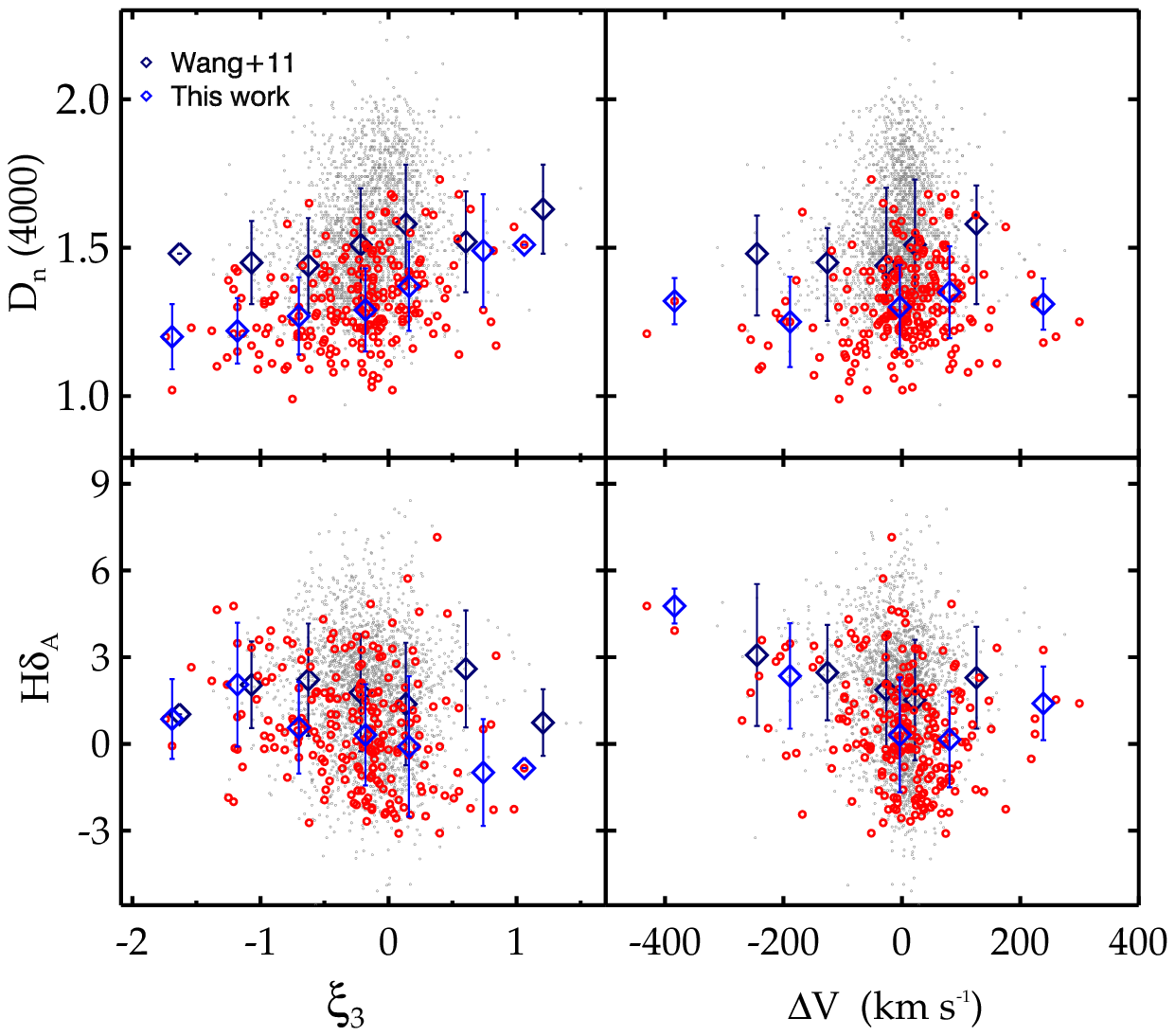}
	\caption{The  Lick indices of \dnfour\ and \hdela\ plotted against the \oiii\ line shape parameters $\xi_3$ (left panels) and the relative bulk velocity shift $\Delta V$ (right panels). 
     The QSO2s and Seyfert 2 galaxies are denoted by the 
    red and grey dots, respectively. In each panel, 
    the median \dnfour\ and \hdela\ values in each bin ($\Delta\xi_3=0.5$ and $\Delta V=100\ \mathrm{km\ s^{-1}}$), 
    along with the uncertainties, are overplotted by the diamonds 
    (dark blue for the Seyfert galaxies and blue for the QSO2s).\rm 
		\label{fig:age}}
\end{figure*}

Both lick indices are plotted against the $\xi_3$ and \oiii$\lambda5007$\ line relative velocity shifts $\Delta\upsilon$ in 
Figure \ref{fig:age} for both QSO2s and Seyfert 2 galaxies\footnote{It is noted that the values of $\delta\upsilon$ in \cite{wang11} are calculated according to the narrow H$\beta$ line for the Seyfert galaxies.}. 
One can see from the figure that,
compared to the Seyfert 2 galaxies,
the QSO2s are biased towards the small \dnfour\ end.
For the QSO2s sample,  about 90\% of the \dnfour\
is less than 1.5, and there is almost no  \dnfour\ above 1.8.  However, a large range extending to 2.0 can be found for 
the \dnfour\ of the Seyfert 2 galaxies. 
The fact that  \dnfour\ increases with stellar population age therefore suggests a young stellar population in QSO2s.
In fact, \dnfour\ = 1.5 is usually used to distinguish young stellar populations from old ones \citep{Kauffmann2003}. Based on the Spearman rank-order test, 
the corresponding correlation coefficients are tabulated in Table \ref{tab:para1}.
For each entry, the value in the bracket is the probability that the two variables are not correlated. In addition to \dnfour, as shown in the figure, the bias against old stellar population in QSO2s can also be learned from the lower two panels
in which the \hdela\ index is used.
\begin{deluxetable}{ccc}
	\tablecaption{Spearman Rank-order Correlation Coefficient Matrix
		\label{tab:para1}
	}
	\tablewidth{0pt}
	\tablehead{
		\colhead{Property} & \colhead{\dnfour}& \colhead{\lledd}\\
		\colhead{} & \colhead{(1)} & \colhead{(2)}  
	}
	\startdata
    \multicolumn{3}{c}{Sample: QSO2s}\\
    \hline
	$\xi_3$ &$0.31 (1.9 \times 10^{-6})$ &$-0.11 (0.10)$ \\
	$\xi_4$ &$-0.09 (0.17) $ &$0.35 (8.6 \times10^{-8}) $\\
    \hline
    \multicolumn{3}{c}{Sample: QSO2s+Seyfert 2s}\\
    \hline
    $\xi_3$ &$0.25 (6.5\times 10^{-24})$ &$-0.21 (2.4\times 10^{-17})$ \\
	$\xi_4$ &$-0.43 (0)  $ &$0.50 (0)$\\
	\enddata
	
\end{deluxetable}
\begin{figure*}[ht!]
	\epsscale{0.9}
	\plotone{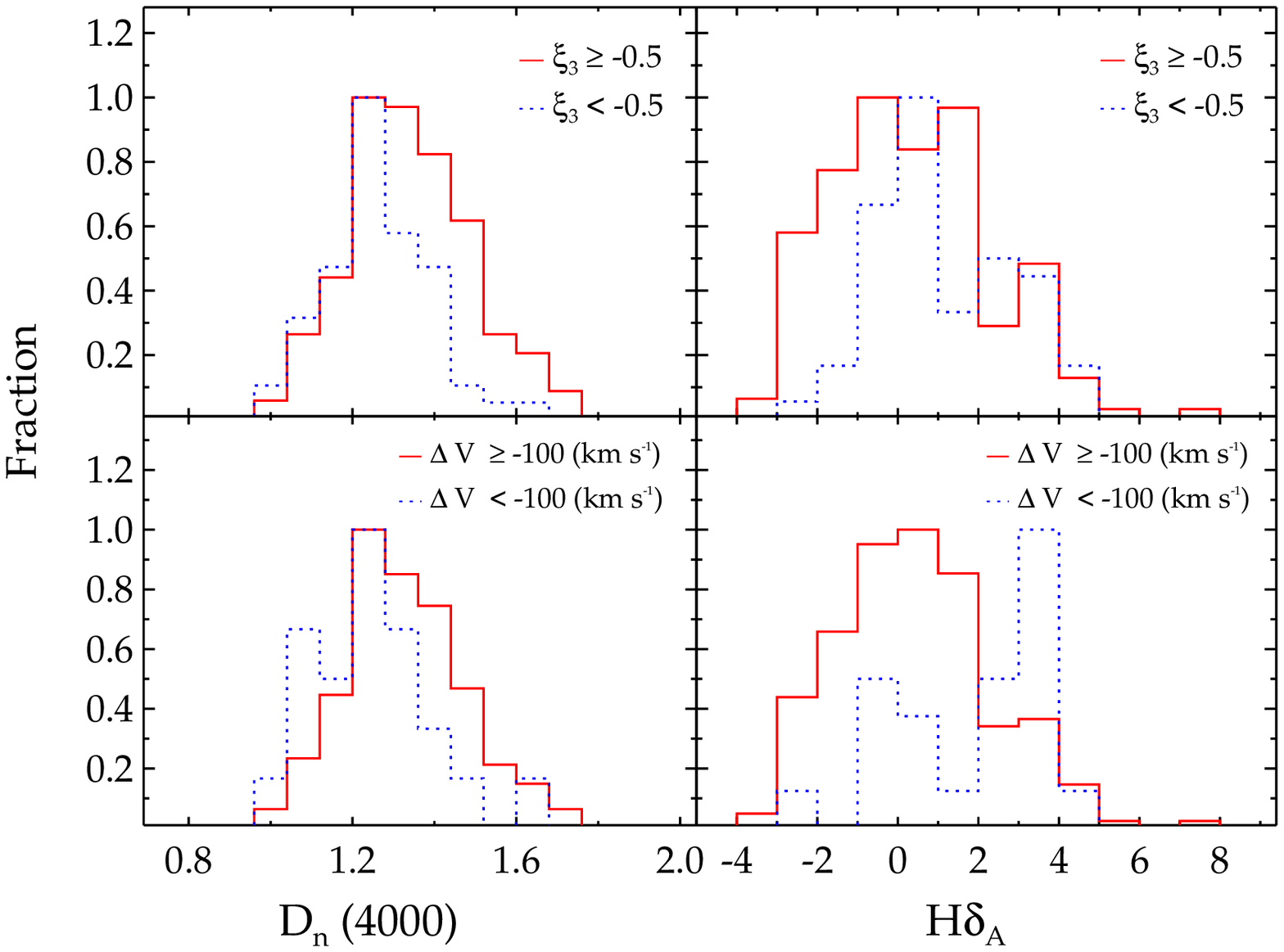}
	\caption{ A comparison of the distribution of \dnfour\ (the left column) and 
\hdela\ (the right column) between the two groups of QSO2s with different shape parameters $\xi_3$ and bulk velocity shifts of the \oiii\ line. 
		\label{fig:fenzu}}
\end{figure*}
To examine the dependence of [\ion{O}{3}] line profile on stellar population,
we divide the QSO2s \rm into two groups, according to their $\xi_3$ values: 
one group has $\xi_3 > -0.5$  and the other has $\xi_3 < -0.5$, by following the method used in \citet{wang11}.  \rm
The corresponding distributions of \dnfour\ and \hdela\ are compared between 
the two groups in the two upper panels in Figure \ref{fig:fenzu}, respectively.  
These plots show that, similar as in the Seyfert 2 galaxies,  the QSO2s with relatively stronger blue asymmetry of the [\ion{O}{3}] line profile tend to be associated with 
younger stellar populations assessed by the smaller \dnfour\ and 
larger \hdela. A similar result can be learned  for $\Delta\upsilon$ from the two lower panels, in which the QSO2s with larger bulk blue velocity shifts tend to have younger stellar populations. 
The difference between the two distributions shown in each panel of Figure 6
is examined by the two-sided K-S tests. 
The calculated maximum distance between the two distributions (see Figure 6 for the details) are tabulated in Table 3. The values in the
brackets are the corresponding significance level at which the two distributions come from the same parent sample.
\begin{deluxetable}{ccc}
	\tablecaption{Matrix of the two-sided K-S tests. 
		\label{tab:para1}
	}
	\tablewidth{0pt}
	\tablehead{
		\colhead{Property} & \colhead{\dnfour}& \colhead{\hdela}\\
		\colhead{} & \colhead{(1)} & \colhead{(2)}  
	}
	\startdata
    \hline
	$\xi_3$ &$0.32 (2\times10^{-4})$ &$0.25 (7\times10^{-3})$ \\
    $\Delta\upsilon$ & $0.29 (6\times10^{-2})$ & $0.45 (4\times10^{-4})$ \\
	\enddata
	
\end{deluxetable} 
\rm 

\subsection{Role of Eddington ratio}

The Eddington ratio $L/L_{\mathrm{Edd}}$ is an important physical parameter driving the AGN's activity. 
\citet{nelson04}  showed a correlation between the \oiii\  $\lambda 5007$ line blue asymmetry and Eigenvector-I space
in the PG quasars.
\citet{shen14} confirmed that \lledd\ is the  main physical driver of the Eigenvector-I space
of AGNs. 
\citet{wang11}  indicated that stronger blue asymmetry is not only correlated with
younger stellar populations but also with higher   \lledd\ for 
a sample of nearby Seyfert  2 galaxies.  
This trend was then confirmed by a sample of nearby partially obscured AGNs. 
\begin{figure*}[ht!]
	\centering
	\epsscale{1.2}
	\plotone{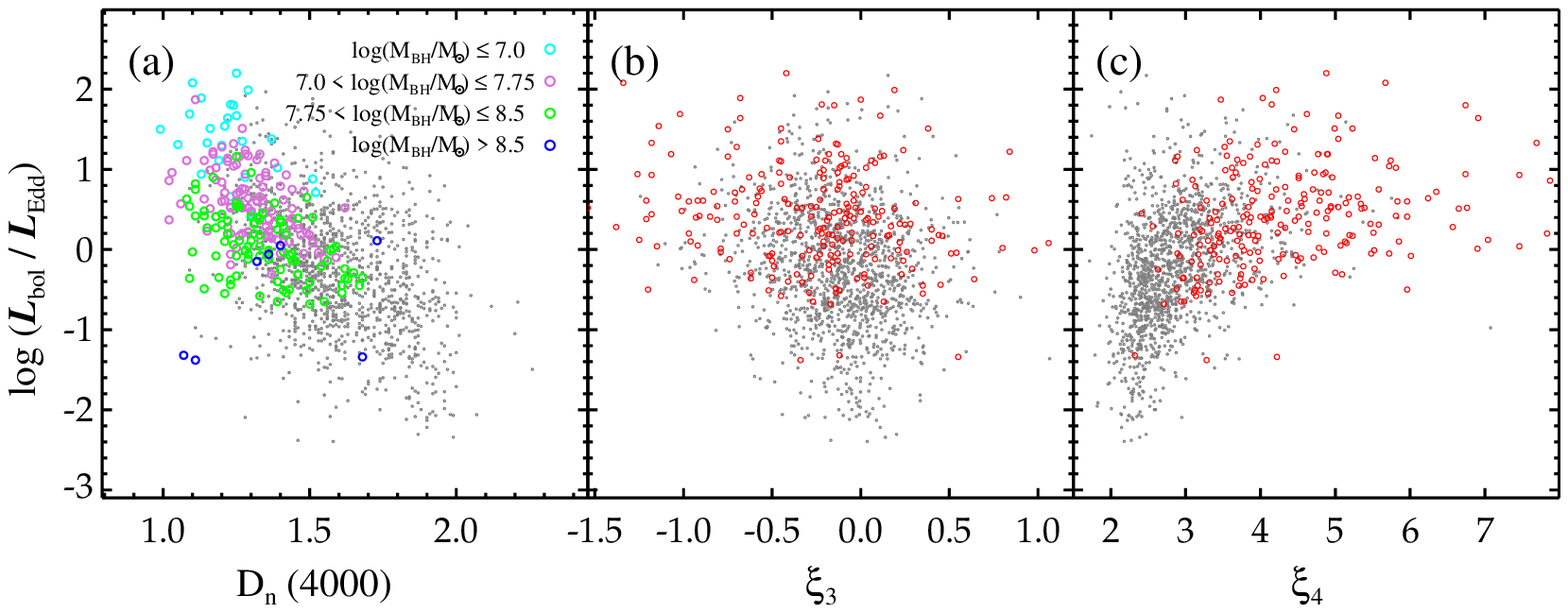}
	\caption{The relationships between \lledd\  and \dnfour\ (left panel), the line profile parameters $\xi_3$ (middle panel), 
	 $\xi_4$ (right panel). The symbols again are the same as in Figure 4. 
		\label{fig:bol}}
\end{figure*}

\subsubsection{Estimation of \texorpdfstring{$L/L_{\mathrm{Edd}}$}. } 
In order to estimate \lledd\ of the QSO2s, the bolometric luminosity  \lbol\ is
 transformed from the intrinsic extinction-corrected  [\ion{O}{3}]$\lambda$5007 line luminosity \loiii\ through
the bolometric correction \lbol$/$\loiii\ $\approx3500$ \citep{heckman04}, which is consistent with that used in \citet{wang11}. 
The extinction-corrected \loiii\ is obtained from \citet{kong18}, in which 
the extinction was estimated from the observed Balmer decrement in the standard case B recombination \citep{halpern83}
and the extinction cure of \citet{cardelli89} with $R_V=3.1$.
A median value of $\rm H_\alpha/H_\beta =4.04$ is adopted for the objects without a measured Balmer decrement.

The black hole mass  $M_{\mathrm{BH}}$ of each object  is estimated from the well-documented $M_{\mathrm{BH}}-\sigma_*$ relationship:
$\log(M_{\mathrm{BH}}/M_\odot)=8.13+4.02\log(\sigma_*/200\ \mathrm{km\ s^{-1}})$ \citep{tremaine02},
in which the stellar velocity dispersion of bulge is replaced by the width of the core of the \oiii\ $\lambda5007$ line.
As presented in \citet{kong18}, the width of the core of the \oiii\ line
can trace the stellar velocity dispersion well for QSO2s.
In addition, a heavy blend between stellar absorption features and  AGN's emission lines,
such as an overlap of \cahk\, [\ion {N}{3}] $\lambda3968$ and $\rm H\epsilon$ lines, leads to a large uncertainty of the measured \sigs.

\subsubsection{Statistics}

Figure \ref{fig:bol} plots \lledd\ as a function of \dnfour, $\xi_3$ and $\xi_4$ from left to right for both QSO2s and Seyfert 2 galaxies sample.
One can see from the left panel that the QSO2s with younger stellar population and higher \lledd\ closely follow the anti-correlation 
between \lledd\ and  \dnfour\ that was previously well established in local AGNs \citep[e.g.,][]{Kewley06,wang11,mullaney13}, which suggests a decrease of \lledd\ as the circumnuclear 
stellar population  continuously ages.  
A Spearman rank-order test yields a correlation coefficient of $ r_s = -0.39 $ with $ P < 10^{-7}$ for the QSO2s.
The correlation coefficient is enhanced to be $ r_s = -0.46 $ with $ P < 10^{-10}$ when the QSO2s and Seyfert 2 galaxies samples
are combined. 

Could the \lledd$-$\dnfour\ anti-correlation be understood by 
an underlying driver due to the mass of the host galaxies (or SMBHs)? 
For example, \citet{Stanley17} indicates that the enhanced star formation
rate in luminous QSOs is resulted from a fact that luminous QSOs tend 
to occur in massive galaxies. To test this alternative, 
we separate the QSO2s into four groups according to their SMBH masses that are believed to be related with the mass of the 
bugle of the host galaxies,
and show the four groups in the panel (a) of Figure 7 by different colors. It is clearly that the \lledd$-$\dnfour\ sequence is still 
valid for the QSO2s with comparable SMBH masses, although there 
is a dependence of the sequence on the SMBH masses.

The relationships between  \lledd\ and the \oiii\ emission line profile parameters $\xi_3$ and $\xi_4$
 are  examined in the  middle  and right panels of Figure  \ref{fig:bol}, respectively. 
Again, the corresponding Spearman rank-order test results are listed in Table \ref{tab:para1}.
Thanks to their high luminosity, an including of the QSO2s therefore reinforces our previous claim
that high \lledd\ is necessary for both strong blue asymmetry and strong broad component of the \oiii\ emission line.
In fact, 
\citet{Zhang2021} recently analyzed the properties of line wing of [\ion{O}{3}] emission line of 535 type I quasars 
by using SDSS optical spectral data, 
and pointed out a dependence of the line wing on the \lledd.

Similar as in \citet {wang11}, correlation between  \lledd\ and 
\delv\  is found neither in the QSO2s sample or a merged sample containing both QSO2s and Seyfert 2 galaxies.   

\section{Discussion}

\subsection{The young stellar population associated with the QSO2s}
We argue that the young stellar populations identified in the QSO2s by the Lick \dnfour\ index is hard to be explained 
by a contamination caused by a underlying AGN's continuum. On the one hand, a two-order 
polynomial, which accounts for a AGN's continuum and an intrinsic extinction, has been involved in our modeling of 
the continuum of the QSO2s in the  {\tt pPXF} package. On the other hand, even without an inclusion of the contribution of the 
AGN's continuum, although a comparison study in \citet {wang15} shows that 
ignoring the AGN's continuum will cause a little underestimation of the measured \dnfour\ value, the level of the underestimation  
has no evident effect on the fact that QSO2s are typically associated with a young stellar population. 

We argue that the fact that the QSO2s are associated with a younger stellar population than the Seyfert galaxies is not due to the aperture effect caused by the fixed 3\arcsec\ fiber width adopted by the SDSS. 
The measured \dnfour\ is plotted as a function of $z$ in  
Figure 8 for a comparison between the QSO2s and the Seyfert 2 galaxies. 
At first, the QSO2s and Seyfert galaxies overlap with each other within a redshift range from 0.05 to 0.15. In this common 
redshift range, the median values of \dnfour\ within each redshift bin of 
0.05 reinforce the conclusion that QSO2s are generally associated with younger stellar populations than Seyfert galaxies do.
Secondly, the stellar population age of the bulge of the high-redshift 
QSO2s is expected to be likely overestimated due to the fixed 
aperture size, because of the radial color gradient of galaxies, 
which is partially resulted from stellar population age 
\citep [e.g.,][and references therein] {Liao23}.

Finally, although being not as strong as in the Seyfert galaxies, 
the value of \dnfour\ sightly decreases with redshift for the QSO2s,   
which is likely due to a cosmic evolution effect. It seems that 
the slight decrease of \dnfour\ with redshift could not be only
explained by a selection effect on luminosity, in which 
high-z QSO2s tend to be more luminous and be associated with younger 
stellar populations. The comparisons in Figure 2, in fact, show that distributions of \oiii\ line luminosity have no  
clear difference between the 221 QSO2s and the parent sample in
both redshift bins.

\begin{figure}[ht!]
	\centering
	\epsscale{1.1}
	\plotone{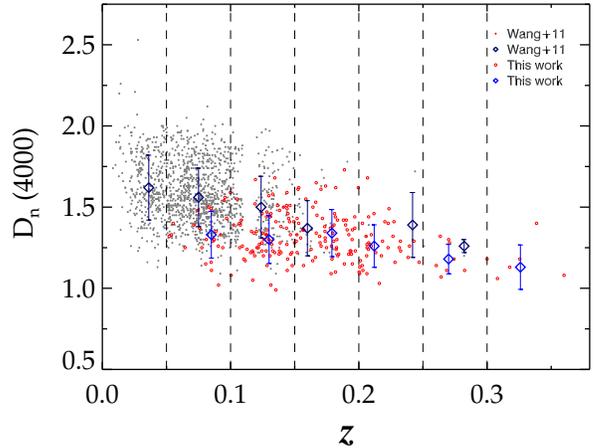}
	\caption{\dnfour\ plotted against redshift for the QSO2s (the red open circles) and Seyfert galaxies (the grey dots). The median values of \dnfour\ determined in each redshift bin of 0.05  are marked by the black and blue diamonds for the QSO2s and Seyfert galaxies , respectively. 
		\label{fig:d4000z}}
\end{figure}
\rm
The young stellar populations in QSO2s have been frequently claimed in previous studies, which implies a quasi-simultaneously triggered central AGN’s activity and circumnuclear starburst \citep [e.g.,][]{Heckman97,Canalizo00,Cana01,Brotherton02,Holt07,Wills08,Liu09,Tadhunter11,Vi12,Bessiere14,Bessiere2017}.
For instance, 
by fitting the spectra of 21 QSO2s through the galaxy stellar populations analysis method, \citet{Bessiere2017} shows that 71\% of 
the QSO2s in the sample contain young stellar populations with the age less than  
the maximum lifetime of 100 Myr \citep [see also in][]{Bessiere14} expected for an AGN  \citep{Martini01}. 
By an identification of the UV absorption features such as \ion{Si}{3}$\lambda$1417 caused by late O and early B supergiants, 
a young stellar population with an age of $\sim$6 Myr has been revealed in the nearby QSO2s Mrk\,447 \citep{Heckman97}.
In addition, young stellar populations with $t<0.1$Gyr are frequently or dominantly found in the quasar-like luminous objects 
with $L_{\mathrm{[OIII]}}>10^{42}\ \mathrm{erg\ s^{-1}}$ \citep [e.g.,][]{Canalizo00,Holt07,Wills08,Tadhunter11}.

\subsection{\texorpdfstring{\lledd}.-driven feedback}

Our study  shows that the outflow from central SMBH in 
QSO2s as 
traced by the \oiii\ line profile (i.e., $\xi_3$ and $\xi_4$)
generally increases with the \lledd\ (see panels b and c in Figure 7),
which suggests a \lledd-driven feedback and is consistent with not only 
the  observational studies in the past decade, but also
the model predictions. 
On the observational ground, there were plentiful studies focusing on the 
relation between the outflow kinematics and the accretion activity of central 
SMBH in past decade for not only Seyfert galaxies, but also their 
luminous cousin, quasars. Briefly speaking, the outflow strength is 
found to increase with SMBH's accretion activity assessed by multiple ways, including  
bolometric luminosity $L_{\mathrm{bol}}$, [\ion{O}{3}]$\lambda5007$ line luminosity, intrinsic hard X-ray luminosity, \lledd, and radio power \citep[e.g.,][]{greeney05,mullaney13,Bae14,Mullaney14,Zakamska14,Woo16,Wang16,kong18,wang18,Davies20}. 
In addition, 
as shown in Figure 7,
the QSO2s show stronger feedback than the Seyfert galaxies due to their higher \lledd.
\rm

On the theoretical ground, the observed outflow is believed to be resulted from the wind/radiation pressure launched from the inner accretion disk
in the wind/radiation model  \citep [e.g.,][]{Murray95,Proga00,Crenshaw03,King03,Pounds03,King05,Ganguly07,Reeves09,Alexander10,Dunn10,King11,Fabian12,Zubovas12}, 
which successfully explained the fast outflows suggested by the blueshifted ultraviolet and X-ray absorption lines (such as \ion{Fe}{15} and \ion{Fe}{16} )\citep [e.g.,][]{Tombesi12,Higginbottom14}. 
Even though the specific launch mechanism is still under debate, 
the extension of the wind launched from the accretion disk can reach at the inner NLR \citep {Proga08}, which is supported by 
recent observations \citep [e.g.,][]{Fischer18,Kang18,Husemann19}. 
In addition, in the merger scenario \citep [e.g.,][]{Di05,Springel05,Hopkins06}, a strong feedback attributed to the AGN activity is required not only to make the quasar activity to be detectable in optics by removing the material enshrouding the central SMBH \citep [e.g.,][]{Hopkins2005},
but also to regulate SMBH growth through quenching the surrounding star formation activity by a feedback \citep [e.g.,][]{Alexander12,Fabian12,Kormendy13}.

\subsection{Feedback in co-evolution of AGNs and their hosts}

Similar as in Seyfert galaxies, we here identify a dependence of 
ionized gas outflow caused by SMBH's accretion on the circumnuclear
stellar population in QSO2s: young stellar population (and also 
high \lledd) is  related to  a strong outflow. 
This result follows the co-evolutionary scenario proposed previously 
\citep [e.g.,][]{wang15} in which 
AGNs likely evolve from a high-\lledd\ state with
strong outflow to a low-\lledd\  state with weak outflow as the newly
formed circumnuclear massive stars fades out continually.
Recent integral-field spectroscopic observations of AGNs in fact
reveal  that  
stronger the outflows, higher star formation rate (SFR) and higher \ion{H}{1} gas fraction will be \citep [e.g., Figures 6 and 9 in][and references therein]{Luo21,Woo20}.

The revealed dependence of outflow on stellar population implies
an evolution of feedback of AGNs with their host galaxies.
However, the feedback effect cause by the outflow is still under hot debate. 
In the evolution scenario proposed in \citet{Sanders1988} ,
a quasar is produced by expelling the surrounding gas and dust by a wind 
from the central SMBH after a merger of two gas-rich galaxies. 
Such feedback from the powerful AGN’s wind is actually involved in the
early numerical and semi-analytical galaxy evolution models to 
reproduce the $M_\mathrm{{BH}}-\sigma_\star$ relation and luminosity 
functions of AGNs by 
quenching the star formation and blowing the gas or dust away, especially 
in the young AGN phase
\citep [e.g.,][]{1999Fabian,Di05,Hopkins2005,Springel05,Croton06,Hopkins08,Hopkins08b,Khalatyan08,Somerville08,Kauffmann09}.

Recent observations, however, 
indicate that the AGN's feedback has positive, negative and even no 
effect on the evolution of the host galaxies of Seyfert galaxies and quasars \citep [e.g.,][and references therein]{Almeida22,Smirnova22}.
By comparing the specific SFR of AGNs with and without outflows, 
\citet{Woo20} proposed a delayed effect of feedback on host galaxies, due to the dynamical time required for outflows to travel the large galaxy disks.
The MUSE integral-field spectroscopic observation of nine nearby
Palomar-Green quasars show that the fraction of kinetic power of 
outflow is $\dot{E}_{\mathrm{kin}}/L_{\mathrm{bol}}\lesssim10^{-3}$ 
\citep [and see also in][]{Baron19,Fiore17,Rojas20,Molina22}
which is much lower than the theoretical requirement of $\dot{E}_{\mathrm{kin}}/L_{\mathrm{bol}}\approx0.05-0.5$ 
\citep [e.g.,][]{Di05,Springel05,Hopkins10,Zubovas12}.
\rm
\section{Summary}

The evolutionary role of the outflow from QSO2s is examined on a large sample of 221 QSO2s extracted from the QSO2s catalog provided in \citet{reyes08}.
Given our spectral analysis on both AGN and its host galaxies, the main results are listed as follows:
\begin{enumerate}
\setlength{\parskip}{0pt}
\item 
Using  the  Lick indices as indicators, QSO2s are confirmed to be 
on average associated with younger stellar populations than do Seyfert galaxies; 

\item 
Even though an occupation at the high \lledd\ end,
the QSO2s follow the \lledd-\dnfour\ sequence established from the local, less-luminous Seyfert galaxies, which suggests a coevolution between the 
accretion activity of SMBH and the host galaxy.

\item 
QSO2s with a stronger outflow and 
higher activity (\lledd) are tend to be associated with a younger 
stellar population, which implies 
a coevolution between the feedback from SMBH and the host in QSO2s driven by \lledd: AGNs likely evolve from a high \lledd state with strong feedback to
a low \lledd state with weak feedback as the circumnuclear stellar population continually ages. 
\end{enumerate} 

\begin{acknowledgments}
The authors thank the anonymous referee for the careful
review and suggestions improving the manuscript significantly.
This work was supported by the National SKA Program of China (Grant No.
2022SKA0120103), the Astronomical Union Foundation under grant No. U1831126 and 
the Natural Science Foundation of Hebei Province No. A2019205100, by
the National Natural Science Foundation of
China (Grants No. 12173009, 12273054 and 12273076). J.W. is supported
by the Natural Science Foundation of Guangxi (2020GXNSFDA238018).

Funding for the SDSS has been provided by the Alfred P. Sloan Foundation, the Participating Institutions, the National Science Foundation, the U.S. Department of Energy, the National Aeronautics and Space Administration, the Japanese Monbukagakusho, the Max Planck Society, and the Higher Education Funding Council for England. The SDSS Web Site is http://www.sdss.org/

The SDSS is managed by the Astrophysical Research Consortium for the Participating Institutions. The Participating Institutions are the American Museum of Natural History, Astrophysical Institute Potsdam, University of Basel, University of Cambridge, Case Western Reserve University, University of Chicago, Drexel University, Fermilab, the Institute for Advanced Study, the Japan Participation Group, Johns Hopkins University, the Joint Institute for Nuclear Astrophysics, the Kavli Institute for Particle Astrophysics and Cosmology, the Korean Scientist Group, the Chinese Academy of Sciences (LAMOST), Los Alamos National Laboratory, the Max-Planck-Institute for Astronomy (MPIA), the Max-Planck-Institute for Astrophysics (MPA), New Mexico State University, Ohio State University, University of Pittsburgh, University of Portsmouth, Princeton University, the United States Naval Observatory, and the University of Washington.
\end{acknowledgments}


\bibliography{mybib}
\bibliographystyle{aasjournal}

\end{document}